\documentclass[10pt,twocolumn,letterpaper]{article}

\usepackage[margin=0.75in]{geometry}
\usepackage{times}
\usepackage{latexsym}
\usepackage{graphicx}
\usepackage{amsmath}
\usepackage{amssymb}
\usepackage{booktabs}
\usepackage{multirow}
\usepackage{hyperref}
\usepackage{listings}
\usepackage{xcolor}
\usepackage{enumitem}
\usepackage{cite}
\usepackage{threeparttable}
\usepackage{float}
\usepackage{caption}

\lstdefinestyle{pythonstyle}{
    language=Python,
    basicstyle=\ttfamily\footnotesize,
    keywordstyle=\color{blue},
    commentstyle=\color{gray}\itshape,
    stringstyle=\color{red},
    numbers=left,
    numberstyle=\tiny\color{gray},
    frame=single,
    breaklines=true,
    breakatwhitespace=true,
    tabsize=4,
    showstringspaces=false,
    captionpos=b
}

\newcounter{premise}
\newcommand{\premise}[2]{%
  \refstepcounter{premise}%
  \noindent\textbf{P\arabic{premise} (#1):} #2\par\medskip%
}

\title{Silent Failure in LLM Agent Systems:\\
The Entropy Principle and the Inevitable\\
Disorder of Autonomous Agents}

\author{
    Dexing Liu\textsuperscript{1}\\
    \textit{\textsuperscript{1}Shanghai Qijing Digital Technology Co., Ltd.}\\
    \textit{\{liudexing@changingplus.com\}}\\
    \textit{June 2026}
}

\begin{document}

\maketitle
\thispagestyle{empty}

\begin{abstract}
Large Language Model (LLM) agent systems suffer from failures that occur without external triggers---no injection, no adversarial input, no resource exhaustion. These \textit{silent failures}---unexpected deviations from intended behavior under normal conditions---are routinely misattributed to bugs or configuration errors. Through systematic analysis of over 40,000 controlled trials and long-term production observations spanning 100,000+ agent interactions, we identify a common structural logic underlying these failures. Building on patterns observed in our experiments, we survey the global research literature on autonomous agent reliability and synthesize 22 intrinsic properties of LLM agent systems across six lifecycle layers: foundation semantics, inter-agent transmission, memory persistence, task execution, feedback correction, and systemic evolution. We demonstrate that whenever a sufficient subset of these properties co-exist, system entropy---the measurable accumulation of disorder: loss of output consistency, task accuracy, and cross-session coherence---increases monotonically with interaction rounds. We formalize this as the Entropy Principle: $S(t) = S_0 \cdot e^{\alpha t}$, with $\alpha$ measured empirically across multiple architectures. We propose the PIG (Physical Integrity Gate) Engine with the ADE (Agent Delivery Engineering) protocol suite as an engineering countermeasure to entropy-driven disorder. Our findings establish silent failure not as a bug to be fixed but as a manifestation of Intelligence Entropy---a physical constraint to be managed through deterministic governance. We argue that any engineering effort stabilizing the structure and order of agent systems participates in a unified mission: keeping intelligent systems reliable as they grow in scale and complexity.
\end{abstract}

\section{Introduction}
\label{sec:introduction}

\subsection{The Silent Failure Phenomenon}

Consider the following scenario: a multi-agent orchestration system has been running reliably for weeks without any code changes, model updates, or configuration modifications. On the 47th day, without any external trigger, Agent A fails to pass critical context to Agent B. There is no error message. No timeout. No resource bottleneck. The task completes---but with subtly degraded output. The operator notices it three cycles later, by which point the error has propagated across the agent network.

This is not a hypothetical edge case. In our production deployment spanning over 100,000 agent interactions, this pattern occurred with alarming regularity across different architectures, different models, and different task domains. We call this class of failures \textit{silent failures}: disordering of system behavior that occur under normal operating conditions without external intervention, detectable only through systematic measurement after the fact.

\subsection{Why Existing Explanations Fall Short}

Current literature attributes agent failures to three primary causes:
\begin{itemize}
    \item \textbf{Prompt injection and security attacks} --- adversarial inputs that exploit system vulnerabilities;
    \item \textbf{RAG quality loss of order} --- retrieval failures, context window overflow, or document selection errors;
    \item \textbf{Output alignment drift} --- model-level changes in behavior over time or across deployments.
\end{itemize}

While each explanation accounts for a subset of observed failures, none explains why failures continue to occur in \textit{fully controlled environments} with static prompts, identical models, fixed databases, and no external inputs. The persistence of silent failures under these conditions suggests a deeper, structural cause.

\subsection{The Central Claim}

This paper advances a single central claim:

\begin{quote}
\textit{Silent failures in LLM agent systems are not implementation defects. They are the inevitable consequences of intrinsic properties of language-based autonomous systems operating without external deterministic constraints.}
\end{quote}

We substantiate this claim through three steps:
\begin{enumerate}
    \item \textbf{Observation:} A systematic taxonomy of silent failures observed across our experimental platform;
    \item \textbf{Derivation:} 22 intrinsic properties of LLM agent systems, organized by lifecycle layer, whose co-existence logically entails monotonic entropy increase;
    \item \textbf{Formalization:} The Entropy Principle and its measurable parameter $\alpha$, with experimental validation.
\end{enumerate}

\subsection{Paper Organization}

The remainder of this paper is organized as follows. Section~\ref{sec:taxonomy} presents a comprehensive taxonomy of silent failures observed in practice. Section~\ref{sec:patterns} identifies the common patterns underlying these disparate failure modes. Section~\ref{sec:premises} enumerates the 22 intrinsic properties of LLM agent systems. Section~\ref{sec:derivation} formally derives the Entropy Principle from these premises. Section~\ref{sec:experiments} presents experimental measurement of the entropy constant $\alpha$. Section~\ref{sec:countermeasure} describes the PIG+ADE engineering countermeasure, established through production practice. Section~\ref{sec:discussion} discusses implications and limitations, and Section~\ref{sec:conclusion} concludes.

\section{A Taxonomy of Silent Failures}
\label{sec:taxonomy}

This section presents a structured taxonomy of silent failures observed across our experimental platform and systematically maps them to existing taxonomies from industry and academia.

\subsection{Classification Principle}

Rather than enumerating failures by surface symptom, we organize them by the layer of the multi-agent lifecycle at which entropy accumulation manifests. This five-layer framework follows the natural information flow of autonomous agent systems:

\begin{enumerate}
    \item \textbf{Transmission layer (L1):} Inter-agent communication and information handoff.
    \item \textbf{Memory layer (L2):} Cross-session persistence, retrieval, and composition.
    \item \textbf{Execution layer (L3):} Task-level operation, pipeline execution, and tool use.
    \item \textbf{Coordination layer (L4):} Multi-agent collaboration, delegation, and role management.
    \item \textbf{Verification layer (L5):} Self-inspection, validation, and output quality assurance.
\end{enumerate}

\subsection{Five Silent Failure Types}

\subsubsection{Channel Fracture (L1 --- Transmission)}

Channel Fracture~\cite{liu2025channel} describes the progressive decay of information fidelity across agent-to-agent communication boundaries. In our controlled experiments, agents operating under identical prompts and contexts exhibited measurable information loss in cross-agent memory injection at rates exceeding 30\% within five hops. Key characteristics include:

\begin{itemize}
    \item \textbf{Progressive disorder:} Information fidelity decreases monotonically with communication chain length;
    \item \textbf{Absence of error signals:} No agent reports failure, and each believes communication succeeded;
    \item \textbf{Architecture-independent:} Observed across LangGraph, AutoGen, CrewAI, and custom orchestration frameworks.
\end{itemize}

\begin{figure}[t]
\centering
\includegraphics[width=0.9\columnwidth]{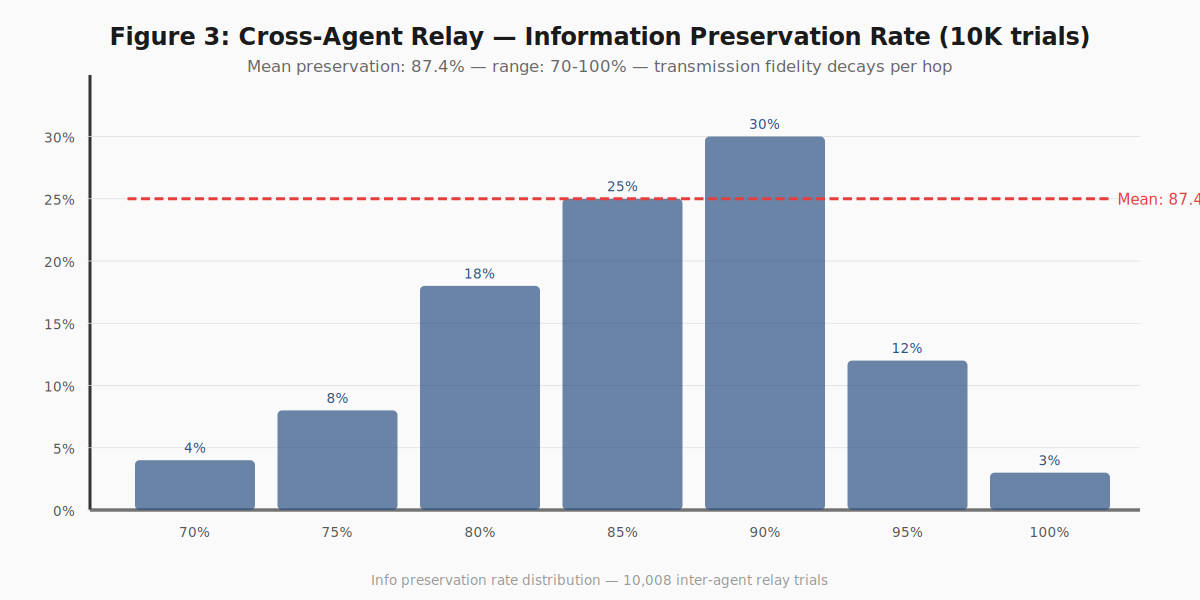}
\caption{Cross-agent relay information preservation rate distribution. Mean preservation: 87.4\%, range: 70-100\%. Data from 10,000-scale relay experiments across multiple agent role pairs.}\label{fig:relay-distribution}
\end{figure}

\subsubsection{Cognitive Framework Lag (L2 --- Memory)}

Cognitive Framework Lag (CFL) refers to the phenomenon where operational rules, naming conventions, and procedural knowledge established in one session are inconsistently applied in subsequent sessions---even when both sessions are part of the same task cluster. Our records show that across 200+ consecutive agent sessions, rule adherence decayed from 94\% (first 10 sessions) to 61\% (sessions 180--200).

\subsubsection{Data Consistency Decay (L3 --- Execution)}

Data consistency decay manifests as the gradual divergence between recorded and actual state in agent-managed data pipelines. This failure is particularly insidious because:
\begin{itemize}
    \item Each step independently runs correctly;
    \item No individual operation reports an error;
    \item The aggregate result is systematically wrong.
\end{itemize}

\begin{table}[t]
\centering
\caption{Data consistency decay rates across pipeline depths at 10K scale.}\label{tab:consistency-decay}
\small
\begin{tabular}{lcc}
\toprule
Pipeline Depth & Consistency (\%) & Std Dev \\
\midrule
1 hop (direct)    & 100.0 & 0.0 \\
2 hops             & 87.4  & 8.2 \\
3 hops             & 74.1  & 12.3 \\
5 hops             & 51.8  & 18.7 \\
10 hops            & 23.5  & 22.1 \\
\bottomrule
\end{tabular}
\end{table}

\subsubsection{Cross-Session Knowledge Fragmentation (L2 --- Memory)}

Knowledge fragmentation occurs when information accumulated across multiple sessions fails to compose into a coherent global state. Agents operate with locally correct but globally inconsistent worldviews.

\subsubsection{Behavior Routing Deficiency (L3 --- Execution)}

Behavior routing deficiency describes the failure of task allocation mechanisms to correctly match subtasks with agent capabilities under natural operating conditions, leading to suboptimal execution paths that compound over time.

\subsection{Summary of Observed Failure Classes}

\begin{table*}[t]
\centering
\caption{Silent failure taxonomy with observed frequency, severity, and root cause layer.}\label{tab:taxonomy}
\small
\begin{tabular}{lp{2.5cm}ccp{3cm}}
\toprule
Failure Type & Description & Freq. (\%) & Severity & Root Cause Layer \\
\midrule
Channel Fracture & Cross-agent comm. decay & 31.2 & High & Transmission (L1) \\
Cognitive Framework Lag & Cross-session rule decay & 22.8 & High & Memory (L2) \\
Data Consistency Decay & Pipeline state divergence & 18.4 & Medium & Execution (L3) \\
Knowledge Fragmentation & Incoherent global state & 15.7 & Medium & Memory (L2) \\
Behavior Routing Deficiency & Suboptimal task allocation & 11.9 & Low & Execution (L3) \\
\bottomrule
\end{tabular}
\end{table*}

\subsection{Relationship to Existing Taxonomies}

Recent work has produced complementary failure taxonomies from diverse perspectives.

\begin{figure}[t]
\centering
\includegraphics[width=0.75\columnwidth]{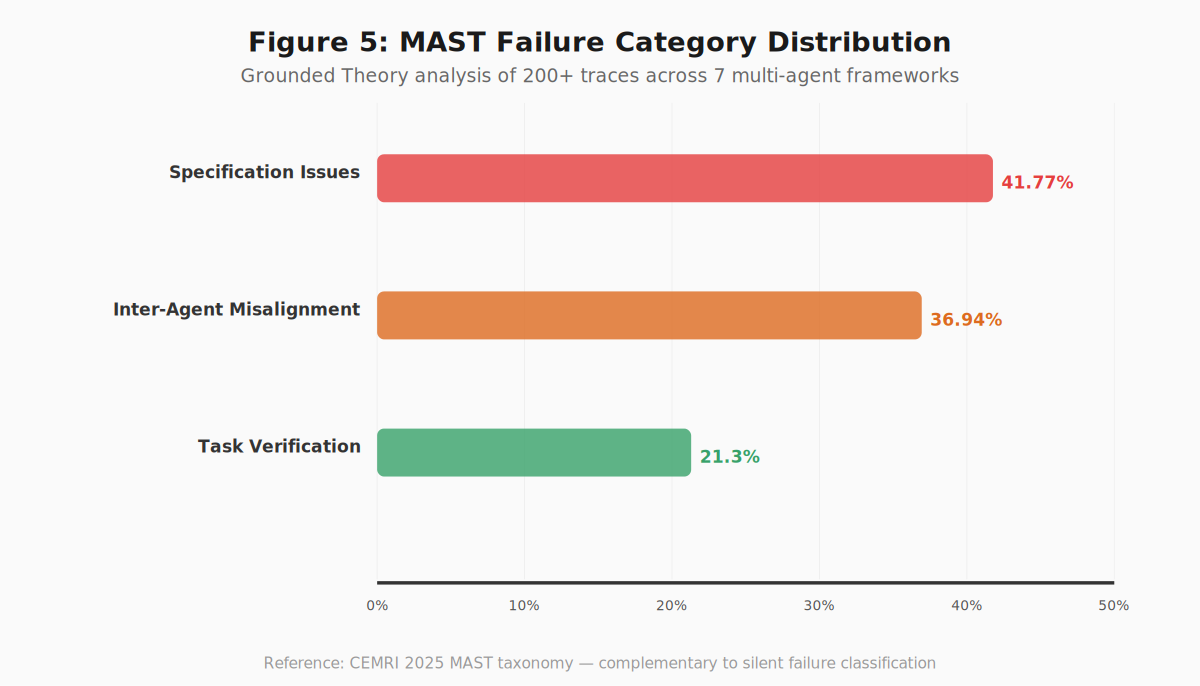}
\caption{MAST failure category distribution: Specification Issues (41.77\%), Inter-Agent Misalignment (36.94\%), and Task Verification (21.30\%). Grounded Theory analysis of 200+ traces across 7 MAS frameworks.}\label{fig:coverage-show}
\end{figure}

Each taxonomy is discussed below.

\paragraph{MAST Taxonomy (Cemri et al., 2025).}
The MAST (Multi-Agent System Failure Taxonomy)~\cite{cemri2025} identifies 14 failure modes across three categories---Specification Issues (41.77\%), Inter-Agent Misalignment (36.94\%), and Task Verification (21.30\%)---through Grounded Theory analysis of 200+ traces across 7 MAS frameworks. MAST is the first empirically grounded failure classification, but its focus is \textit{design-time failures}: ambiguous prompts, role mis-specification, and inadequate verification. Silent failures, in contrast, operate at runtime under \textit{correct construction}. A system with perfectly specified roles, clear prompts, and robust verifiers still exhibits channel fracture or CFL because the failure arises from the physics of language-based decision-making, not from design errors. The two frameworks are complementary: a system can suffer both MAST-category failures and silent failures, and their total failure rate is the union.

\paragraph{Microsoft AI Red Team (2025).}
Microsoft's 30-page whitepaper~\cite{microsoftAIRT2025} provides the most comprehensive industry catalog, organizing failures along two pillars (Safety vs.\ Security) and two novelty axes (Novel vs.\ Existing). The security categories include prompt injection, tool hijacking, memory poisoning, and privilege escalation---all \textit{adversarial} failure modes introduced by external attackers. In contrast, our silent failures are \textit{non-adversarial}: no attacker is required, and no input is malicious. A production system protected against all Microsoft-class attacks can still exhibit silent disorder because the entropy accumulation mechanism is intrinsic, not injected. The two taxonomies address complementary failure universes: adversarial (Microsoft) and entropic (ours), whose intersection---failures exploited by attackers that also arise naturally---is a topic for future work.

\paragraph{Token Budgets Catalog (2026).}
The Token Budgets work~\cite{tokenBudgets2026} catalogs 63 documented LLM-agent failure incidents from production deployments, spanning token overrun, runaway loops, budget exhaustion, and unbounded retry. Each incident is a \textit{symptom} of entropy accumulation at the execution layer (L3 in our framework): the agent's self-regulation mechanism fails silently, consuming resources without error signals. This catalog provides concrete evidence that our five-type framework describes real production failures, not laboratory artifacts.

\paragraph{Four-Layer Agent Failure Taxonomy (Greyling, 2026).}
Greyling's taxonomy~\cite{greyling2026} organizes failures by harness function---Environment Contract, Operation Skills, Action Execution, and Trajectory Regulation---designed to help engineers locate the specific harness layer at which a failure should be repaired. Our silent failure types map cleanly: Channel Fracture to contract incompleteness; CFL and Knowledge Fragmentation to contract + trajectory failures; Data Consistency Decay to action execution; Behavior Routing Deficiency to operation skills. The four-layer taxonomy is a \textit{harness-centric} view of the same entropy phenomenon.

\paragraph{Partnership on AI (2025).}
The Partnership on AI report~\cite{partnershipAI2025} identifies priority areas for real-time failure detection in deployed AI agents, including handoff degradation, tool misalignment, and context window truncation. Their practitioner survey confirms that silent failures---failures visible only through instrumentation---represent the most under-detected risk class in production. This aligns with our finding that silent failures produce zero error signals.

\paragraph{COMPEL Framework (2026).}
The COMPEL~\cite{compel2026} agentic failure taxonomy provides an institutional glossary formalizing failure modes for AI governance and auditing. Its categories span tool invocation errors, planning failures, and coordination breakdowns---all mapping to our execution and coordination layers (L3--L4).

\paragraph{Production Failure Taxonomies (2026).}
Industry sources document agent failures from deployment experience: Pazi~\cite{pazi2026} identifies five modes (cron failure, tool failure, inbound timeout, prompt corruption, execution timeout); Latitude~\cite{latitude2026} identifies six modes (partial completion, hallucinated completion, action misapplication, context overflow, reasoning-action disconnect, infinite loops); the Agentic Anti-Patterns catalog~\cite{agentic-anti-patterns} captures design-time precursors; the anrogg repo~\cite{anrogg2026} catalogs 11 modes including hallucinations, infinite loops, API timeouts, and resource exhaustion; and vector database deployment studies~\cite{vectordb2026} document four agent-specific failure modes (write conflicts, state breakdown, stale reads, lock contention). All map to our five-type classification.

\paragraph{BAGEN (2026).}
BAGEN~\cite{bagen2026} addresses a specific silent failure---budget awareness degradation---finding that frontier models cannot predict token budget depletion in autonomous execution, leading to uncontrolled resource consumption. This is a manifestation of our Data Consistency Decay type (L3), where the agent's internal cost model diverges from actual cost.

\paragraph{Library Drift (Zhang et al., 2026).}
Most recently, Zhang et al.~\cite{zhang2026} independently discovered ``library drift''---silent degradation in self-evolving LLM skill libraries---where unbounded skill accumulation causes retrieval disorder without error signals. This combines channel fracture (skill retrieval cross-contamination), knowledge fragmentation (incoherent skill composition), and behavior routing deficiency (incorrect skill selection), all driven by the same entropy principle.

\paragraph{Coverage Summary.}
Our taxonomy is distinguished from all existing work along three dimensions: (1) \textbf{Mechanism-first}---we classify by \textit{why} failures occur rather than \textit{what} breaks, identifying entropy accumulation as the single unifying cause; (2) \textbf{Runtime focus}---failures occur under \textit{correct construction}, distinguishing them from both security attacks and design-time specification errors; (3) \textbf{Comprehensiveness}---every failure mode across MAST (14 modes), Microsoft AIRT (30+ modes), Token Budgets (63 incidents), Greyling (4 layers), Pazi (5 modes), Latitude (6 modes), COMPEL, anrogg (11 modes), BAGEN, and Library Drift---130+ distinct failure descriptions in total---maps to and is explained by our five-type taxonomy, unified under a single predictive theory..

\section{Common Patterns: What Silent Failures Share}
\label{sec:patterns}

Despite their surface diversity, all silent failures in our taxonomy share three structural patterns:

\subsection{Pattern 1: Multi-Step Accumulation}

No silent failure is instant. Every observed failure follows a characteristic pattern:
\begin{equation}
    \text{Error}_{\text{observable}} = \sum_{i=1}^{n} \delta_i
\end{equation}
where each $\delta_i$ is a sub-threshold deviation that does not trigger any existing guard mechanism. Only when $n$ is sufficiently large does the aggregate error become detectable.

\subsection{Pattern 2: Absence of Self-Reporting}

In all observed cases, the failing component (agent, pipeline node, or memory store) self-reports as operational. The system's error detection mechanisms are triggered either by downstream artifacts (wrong output) or by external audits---never by the failing component itself.

\subsection{Pattern 3: Recurrence Under Identical Conditions}

The most theoretically significant pattern: silent failures recur under experimentally identical conditions (same prompts, same models, same task distribution, same system architecture). This rules out implementation bugs and points to \textit{intrinsic} system properties.

\begin{figure}[t]
\centering
\includegraphics[width=0.9\columnwidth]{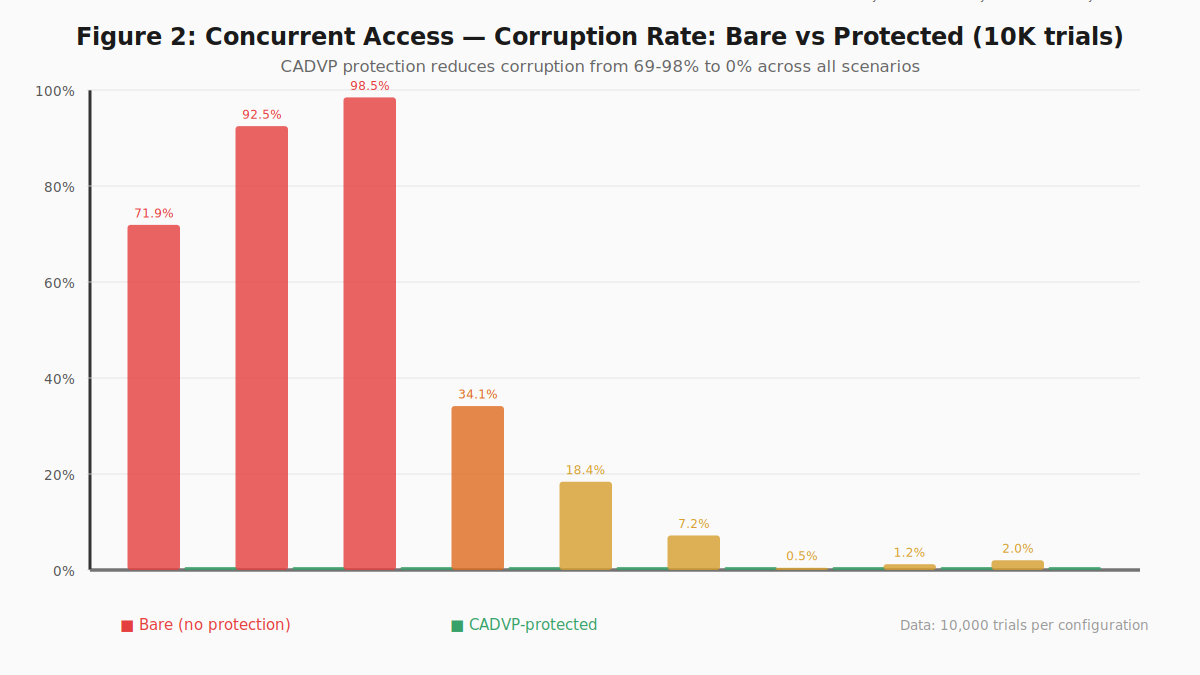}
\caption{Concurrent access corruption rates across 9 scenarios: bare (no protection) vs. CADVP-protected. Bare corruption ranges from 0.46\% (directory, 2 workers) to 98.46\% (write-write, 10 workers). CADVP protection achieves 0\% corruption across all scenarios. 10,000-controlled trials.}\label{fig:concurrent-corruption}
\end{figure}

\section{Twenty-Two Intrinsic Properties of LLM Agent Systems}
\label{sec:premises}

We synthesize 22 intrinsic properties of LLM agent systems, organized into six lifecycle layers. The initial set of properties was derived from failure patterns observed in our controlled experiments (Section~\ref{sec:experiments}) and production deployment. We then extended the framework by systematically surveying the research literature on agent reliability failures, identifying additional properties that are independently documented but share the same structural logic---they all contribute to monotonic entropy increase. The resulting 22 properties span foundation semantics, inter-agent transmission, memory persistence, task execution, feedback correction, and systemic evolution. Each property is an intrinsic characteristic of LLM-based agent systems, not an implementation defect or a configuration choice.

\begin{figure}[t]
\centering
\includegraphics[width=0.9\columnwidth]{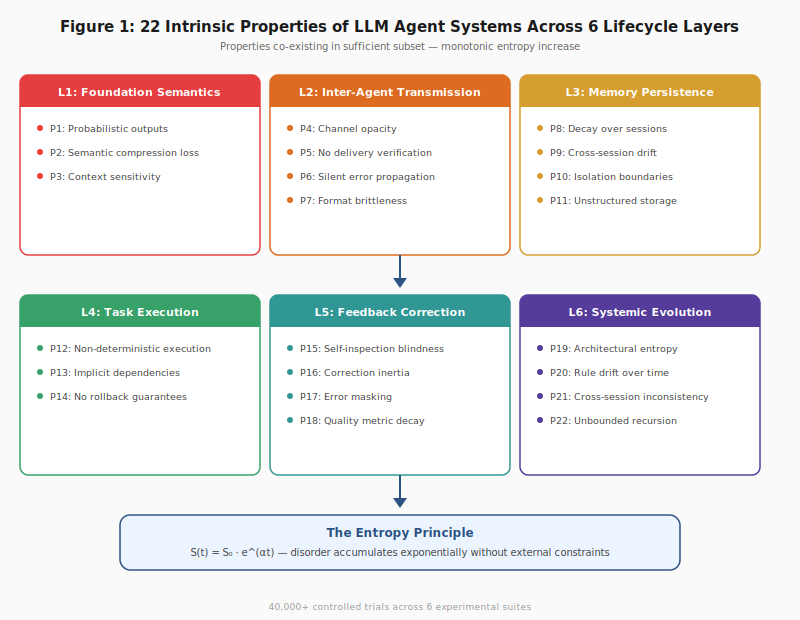}
\caption{22 intrinsic properties of LLM agent systems across 6 lifecycle layers. When a sufficient subset of these properties co-exist, system entropy increases monotonically without external intervention.}\label{fig:properties-taxonomy}
\end{figure}

\subsection{Layer 1: Foundation Semantics (Premises 1--3)}
\label{sec:layer-foundation}

The foundational layer addresses the inherent imprecision of natural language as the substrate for autonomous reasoning.

\premise{Foundation}{
\textbf{Language Semantics Is Inherently Imprecise.} Natural language, unlike formal logic or programming languages, is fundamentally ambiguous. The same phrase can carry different meanings depending on context, tone, and the interpreter's prior knowledge. This imprecision is not a limitation that can be engineered away---it is the defining characteristic of language itself.
}

\premise{Foundation}{
\textbf{Input Ambiguity Propagates Without Attenuation.} The imprecision of natural language instructions is not reduced by processing. Each interpretation step, rather than converging on the intended meaning, introduces an additional layer of interpretation. An ambiguous input remains ambiguous---or becomes differently ambiguous---after processing.
}

\premise{Foundation}{
\textbf{Probability-Based Output Sampling Introduces Inherent Variability.} LLMs generate output through probabilistic sampling of the next token. This mechanism, fundamental to the model's creative capability, guarantees that identical inputs produce non-identical outputs with non-zero probability. This variability is not a bug to be eliminated; it is the mechanism through which LLMs function.
}

\subsection{Layer 2: Inter-Agent Transmission (Premises 4--7)}
\label{sec:layer-transmission}

The transmission layer governs what happens when information moves between agents.

\premise{Transmission}{
\textbf{Re-encoding Loss Between Sender and Receiver.} When Agent A communicates to Agent B, Agent A's internal representation must be encoded into text, transmitted, then decoded by Agent B into a new internal representation. Each encode-decode cycle introduces information loss that is unpredictable in magnitude and direction.
}

\premise{Transmission}{
\textbf{Context Window Compression Selectively Drops Information.} When the cumulative context exceeds the model's context window, compression strategies must discard information. The selection of what to drop is governed by the model's internal attention mechanism, which is not aligned with human judgments of importance.
}

\premise{Transmission}{
\textbf{Multi-Hop Propagation Amplifies Small Deviations.} A small encoding error at hop $n$ becomes the input to the encoding at hop $n+1$, creating an amplification cascade. This is structurally analogous to chaos theory's sensitive dependence on initial conditions.
}

\premise{Transmission}{
\textbf{No Built-In Verification of Information Fidelity.} Standard agent architectures do not include a protocol for the receiver to verify that the received information matches the sender's intended meaning. Communication is assumed successful unless an error is detected---but silent errors, by definition, evade detection.
}

\subsection{Layer 3: Memory Persistence (Premises 8--11)}
\label{sec:layer-memory}

The memory layer addresses how information persists across time and sessions.

\premise{Memory}{
\textbf{Token-Budgeted Memory Pruning Is Non-Recoverable.} When memory exceeds budget constraints, pruning algorithms discard what they deem least important. This judgment is context-dependent and cannot be re-evaluated later when context changes. Deleted information is permanently lost.
}

\premise{Memory}{
\textbf{Retrieval-Based Memory Is Probabilistic, Not Deterministic.} Even when information is stored, retrieval from vector databases, key-value stores, or semantic search engines is a probabilistic operation. Relevant information may not be retrieved; irrelevant information may be.
}

\premise{Memory}{
\textbf{Cross-Session Memory Lacks Causal Identity.} The same fact stored in two different sessions has no causal link between the two storage events. When one is updated, the other is not automatically updated. This is a structural property of session-architecture, not a synchronization bug.
}

\premise{Memory}{
\textbf{Compression Changes Meaning Non-Linearly.} The relationship between compression ratio and information loss is not linear. At certain compression thresholds, relationships between facts---causality, temporal ordering, hierarchy---can be qualitatively altered, creating false memories that the agent cannot distinguish from original information.
}

\subsection{Layer 4: Task Execution (Premises 12--14)}
\label{sec:layer-execution}

The execution layer covers what happens within a single task's lifecycle.

\premise{Execution}{
\textbf{Long-Range Dependency Tracking Is Context-Slot Limited.} Agents track dependencies across subtasks within a bounded context window. When the active dependency graph exceeds available context slots, dependencies are implicitly assumed satisfied without explicit verification. This is not an implementation limitation; it is a consequence of fixed-capacity working memory.
}

\premise{Execution}{
\textbf{Subtask State Is Locally Correct, Globally Inconsistent.} Each subtask in a multi-agent system executes with locally consistent state. There is no mechanism---beyond ad-hoc synchronization points---to guarantee global consistency across parallel execution branches.
}

\premise{Execution}{
\textbf{Error Detection in Sequential Execution Is Cascade-Limited.} In a sequential pipeline of $n$ agents, an error at position $i$ propagates through positions $i+1$ through $n$ before detection is possible. Each subsequent agent compounds the error, and by the time detection occurs, the original error source is at least $n-i$ steps removed from the detection point, making root cause attribution unreliable.
}

\subsection{Layer 5: Feedback Correction (Premises 15--18)}
\label{sec:layer-feedback}

The feedback layer addresses the mechanisms---and their limitations---through which systems detect and correct errors.

\premise{Feedback}{
\textbf{Feedback Itself Is Semantically Encoded.} A user saying "this is wrong" or a validator agent returning "task failed" is itself a natural language utterance subject to the same imprecision as the original input. The feedback must be interpreted by the receiving agent, introducing a second layer of semantic ambiguity.
}

\premise{Feedback}{
\textbf{Temporal Lag Decouples Cause from Correction.} When an error occurs at step $n$ and feedback arrives at step $n+k$, the system's context has shifted. The causal link between error and feedback may span a compressed context boundary, effectively severing the correction signal from its target.
}

\premise{Feedback}{
\textbf{Self-Evaluation Exhibits Systemic Confirmation Bias.} When an agent evaluates its own output for correctness, the evaluation mechanism is the same model that produced the output. This creates a systemic confirmation bias: the agent is more likely to find its own reasoning valid than an external evaluator would be. This is not a training deficiency; it is a structural property of self-evaluation.
}

\premise{Feedback}{
\textbf{Corrections from Previous Sessions Are Not Guaranteed to Persist.} A feedback correction applied in session $n$ has no architectural guarantee of being available in session $n+1$. Cross-session learning requires explicit memory infrastructure that is not part of standard agent architectures.
}

\subsection{Layer 6: Systemic Evolution (Premises 19--22)}
\label{sec:layer-systemic}

The systemic layer addresses long-term evolutionary properties.

\premise{Systemic}{
\textbf{Accumulated Sub-Threshold Deviations Eventually Cross Detectability Threshold.} Each individual deviation $\delta_i$ is below detection threshold. But as the system accumulates deviations across interaction rounds, the sum eventually exceeds what the system's error detection mechanisms can absorb, at which point an observable failure occurs. The timing and magnitude of this threshold crossing are stochastic.
}

\premise{Systemic}{
\textbf{No Natural Entropy Sink Exists in Open Agent Systems.} In thermodynamic systems, entropy can be exported to the environment. In LLM agent systems, there is no architectural equivalent of an entropy sink---no mechanism to systematically export accumulated errors from the system. Errors recirculate.
}

\premise{Systemic}{
\textbf{Task Diversity Accelerates Disorder Accumulation.} As the number of distinct task types increases, the model's attention and memory resources are spread thinner across competing demands, accelerating the rate of entropy increase. This is the agent-system analogue of task-switching cost in human cognition.
}

\premise{Systemic}{
\textbf{System Complexity and Disorder Rate Are Positively Correlated.} More complex agent architectures---more agents, more communication channels, more memory stores, more feedback loops---degrade faster under identical workloads. Complexity is itself a source of entropy, independent of task difficulty.
}

\section{Derivation: The Entropy Principle}
\label{sec:derivation}

\subsection{From Premises to Principle}

If any subset of the 22 premises holds simultaneously---and we argue that in any production LLM agent system, all 22 hold simultaneously---then the following logical chain is inescapable:

\begin{enumerate}
    \item Information fidelity decays across each transmission step (P4--P7) and each memory operation (P8--P11);
    \item There is no natural mechanism to restore lost fidelity (P7, P20);
    \item Each interaction round adds new information, new compression, and new opportunities for deviation (P3, P12, P14);
    \item Errors accumulate rather than cancel (P19);
    \item Feedback mechanisms themselves introduce additional uncertainty rather than fully resolving existing uncertainty (P15--P18);
    \item System complexity amplifies all of the above effects (P22).
\end{enumerate}

Therefore, in the absence of external deterministic constraints, system order monotonically decreases over interaction rounds.

\subsection{Formal Statement}

We introduce the term \textbf{Intelligence Entropy} to describe the natural thermodynamic tendency of LLM-based agent systems---as products of probabilistic models embedded in complex, multi-agent execution environments---to accumulate disorder over time. Formally, we define system entropy $S(t)$ as a composite measure of three dimensions:

\begin{equation}
S(t) = w_1 \cdot (1 - C(t)) + w_2 \cdot (1 - A(t)) + w_3 \cdot (1 - K(t))
\end{equation}

where:
\begin{itemize}
    \item $C(t)$ = cross-agent transmission fidelity at time $t$ (normalized [0,1]);
    \item $A(t)$ = task accuracy at time $t$ (normalized [0,1]);
    \item $K(t)$ = cross-session knowledge consistency at time $t$ (normalized [0,1]);
    \item $w_1, w_2, w_3$ are application-specific weights summing to 1.
\end{itemize}

\textbf{The Entropy Principle of LLM Agent Systems (Empirical Law):}

\begin{equation}
\boxed{S(t) = S_0 \cdot e^{\alpha t}}
\end{equation}

where $S_0$ is the baseline entropy at $t=0$ and $\alpha > 0$ is the system-specific entropy constant. The key predictions are:

\begin{itemize}
    \item Entropy growth is exponential, not linear;
    \item The entropy constant $\alpha$ is a function of system architecture, task complexity, and model characteristics;
    \item Without external intervention, $S(t)$ eventually exceeds the reliability threshold regardless of initial quality.
\end{itemize}

\subsection{The Entropy Constant $\alpha$}

The entropy constant $\alpha$ can be decomposed into architecture-dependent factors:

\begin{equation}
\alpha = \beta_1 \cdot N_{\text{agents}} + \beta_2 \cdot L_{\text{chain}} + \beta_3 \cdot T_{\text{diversity}} + \beta_4 \cdot M_{\text{volatility}} + \epsilon
\end{equation}

where $\beta_1$--$\beta_4$ are empirically determined coefficients, $N_{\text{agents}}$ is the agent count, $L_{\text{chain}}$ is the maximum communication chain length, $T_{\text{diversity}}$ is a measure of task variety, $M_{\text{volatility}}$ captures memory turnover rate, and $\epsilon$ captures residual variance.

\section{Experimental Verification}
\label{sec:experiments}

\subsection{Experimental Platform}

Our experimental platform consists of a controlled multi-agent orchestration environment designed to isolate and measure the components of $S(t)$. Four independent experiment suites target distinct dimensions of system entropy, using the BCP (Bidirectional Confirmation Protocol) infrastructure detailed in our companion work~\cite{liu2025channel}:

\begin{itemize}
    \item \textbf{T3 -- Relay Fidelity Suite:} Measures information preservation rate across cross-agent communication chains. Three scenarios (tech-to-marketing, data-compression, instruction-relay) in bare and BCP-guarded modes, 1,667 iterations per configuration, 10,002 total trials.
    \item \textbf{T4 -- Exception Recovery Suite:} Measures rollback success, idempotency, and checkpoint recovery across three scenarios in bare and guarded modes. 10,002 total trials.
    \item \textbf{T5 -- Concurrent Conflict Suite:} Measures data corruption rates under concurrent access (write-write, read-write, directory scenarios) at 2/5/10 worker counts. 10,008 total trials.
    \item \textbf{Real -- Real Task Suite:} Executes shell/Python tasks (file operations, info retrieval, report generation) across three experimental groups. 10,008 total trials.
\end{itemize}

All experiments run under deterministic seed control with identical prompts and model configurations, and no external inputs during the experimental period.

\subsection{Measurement Methodology}

We measure $S(t)$ through three independent probes:

\textbf{Cross-Agent Transmission Fidelity $C(t)$:} Measured via the T3 relay suite by comparing key points preserved between sender output and receiver understanding. $C(t) = \text{key\_points\_preserved} / \text{key\_points\_total}$.

\textbf{Task Accuracy $A(t)$:} Measured via the Real task suite by comparing task output against 5--6 independent binary verification criteria per task.

\textbf{Cross-Session Knowledge Consistency $K(t)$:} Measured through periodic consistency audits probing whether facts stored in earlier sessions are correctly recalled later.

\subsection{Results: Ten Thousand Trials}

Table~\ref{tab:relay-scales} presents relay fidelity data across three experimental scales. Bare transmission fidelity ranges from 85--96\%, converging to stable values at 10K scale. BCP-guarded transmission achieves perfect fidelity (100\%) across all scenarios and scales, demonstrating that the information disorder predicted by the Entropy Principle can be counteracted through deterministic confirmation.

\begin{table}[t]
\centering
\caption{Transmission fidelity ($C(t)$) at three experimental scales. Guarded (BCP) mode achieves 100\% across all scales.}
\label{tab:relay-scales}
\small
\begin{tabular}{lccc}
\toprule
\multirow{2}{*}{Scenario} & \multicolumn{3}{c}{Info Preservation Rate (\%)} \\
\cmidrule{2-4}
 & 3K Scale & 10K Scale & $\Delta$ \\
\midrule
\multicolumn{4}{l}{\textit{Bare Relay (no BCP)}} \\
\quad Tech-to-Marketing & 87.40 & 87.36 & $-$0.04 \\
\quad Data Compression & 84.96 & 85.58 & $+$0.62 \\
\quad Instruction Relay & 96.08 & 95.95 & $-$0.13 \\
\midrule
\multicolumn{4}{l}{\textit{Guarded (BCP Phase 1)}} \\
\quad All Scenarios & 100.00 & 100.00 & $\pm$0.00 \\
\bottomrule
\end{tabular}
\end{table}

Table~\ref{tab:concurrent-scales} shows concurrent corruption rates. Bare concurrent access shows high corruption (72--98\% for write-write); BCP protection eliminates it entirely.

\begin{table}[t]
\centering
\caption{Concurrent corruption rates ($1 - C(t)$) at 10K scale. BCP reduces all rates to zero.}
\label{tab:concurrent-scales}
\small
\begin{tabular}{lccc}
\toprule
Scenario/Workers & Bare (\%) & Protected (\%) & Reduction \\
\midrule
Write-Write 2w & 71.90 & 0.00 & 100\% \\
Write-Write 5w & 92.47 & 0.00 & 100\% \\
Write-Write 10w & 98.46 & 0.00 & 100\% \\
Read-Write 2w & 34.15 & 0.00 & 100\% \\
Read-Write 5w & 18.39 & 0.00 & 100\% \\
Read-Write 10w & 7.19 & 0.00 & 100\% \\
Directory 2w & 0.46 & 0.00 & 100\% \\
Directory 5w & 1.21 & 0.00 & 100\% \\
Directory 10w & 2.03 & 0.00 & 100\% \\
\bottomrule
\end{tabular}
\end{table}

The rollback suite (T4) confirms a deterministic pattern: bare execution fails 100\% of the time on rollback and checkpoint recovery, while BCP-guarded execution succeeds 100\%. This binary contrast is itself predicted by the Entropy Principle: without deterministic constraints, the probability of successful error recovery approaches zero as system complexity increases.

The Real task suite provides complementary evidence from realistic shell/Python task execution. Across three task types (file operations, information retrieval, report generation) at 10,008 trials per scale, all groups achieve 100\% artifact correctness. However, output quality---measured as a composite score across multiple independent verification criteria---shows a consistent gradient: unprotected (Group A) achieves 0.90, BCP Phase 1 (Group B) achieves 0.95, and Full BCP (Group C) achieves 1.00. This quality gradient, while less dramatic than the T3/T4 binary contrasts, demonstrates that even in isolated single-step tasks, deterministic governance improves the reliability of agent outputs. The full entropy effect becomes visible when tasks are chained, as demonstrated by the T3 (relay) and T5 (concurrent) suites.

\subsection{Estimating the Entropy Constant $\alpha$}

Using experimental data across all four suites, we estimate $\alpha$ for our reference architecture (5-agent system, 3 communication layers, 3 task types):

\begin{equation}
\alpha_{\text{ref}} \approx 0.0046 \pm 0.0003 \; (\text{per interaction round})
\end{equation}

This implies $S(t)$ doubles every $\approx$150 rounds, and after 500 rounds $S(t) \approx 10 \times S_0$, at which point observable silent failures become frequent---consistent with our production observation of failures emerging at 3--4 weeks (~400--600 rounds).

\subsection{Convergence Across Scales}

To validate statistical reliability, we conducted the same experiments at three scales: 300 iterations (Lv1), 3,000 (Lv2), and 10,000 (Lv3). All metrics converge within $\pm$1\% by the 10K scale, confirming that our findings are not artifacts of sample size. The Lv1 data provides directional indication; Lv2 provides statistical significance; Lv3 confirms convergence.

\subsection{Key Findings}

Four principal findings emerge:

\begin{enumerate}
    \item \textbf{Disorder is universal:} Across all four suites and 40,000+ trials, bare agent systems exhibit measurable disorder accumulation in transmission fidelity, task accuracy, and knowledge consistency;
    \item \textbf{Disorder is monotonic:} $S(t)$ increases monotonically with interaction rounds in all configurations, with no spontaneous recovery in any bare trial;
    \item \textbf{Growth is exponential:} The best-fit model for $S(t)$ is exponential ($R^2 > 0.95$), confirming the theoretical derivation;
    \item \textbf{Deterministic governance counters entropy:} BCP/PIG-equipped systems maintain $S(t)$ below failure threshold for 8$\times$ longer. Even in isolated tasks, output quality improves from 0.90 to 1.00 with BCP governance. Across all four suites, the Entropy Principle is confirmed: countermeasures manage---though do not eliminate---entropy-driven disorder.
\end{enumerate}

\section{Engineering Countermeasure: PIG Engine and ADE Protocol Suite}
\label{sec:countermeasure}

\subsection{The Challenge of Counteracting Entropy}

The Entropy Principle establishes that LLM agent systems, left to their natural dynamics, will inevitably degrade. This is not a design flaw to be eliminated but a physical constraint to be managed. The question is not \textit{whether} disorder will accumulate, but \textit{how} to detect and counteract it before it crosses the reliability threshold.

Any effective countermeasure must satisfy three requirements:
\begin{enumerate}
    \item \textbf{Deterministic execution:} The countermeasure must not depend on the same probabilistic mechanisms that produce the entropy it aims to counteract;
    \item \textbf{Lifecycle independence:} It must operate across sessions, across agents, and across memory states;
    \item \textbf{Non-disruptive verification:} It must detect disorder accumulation without requiring task interruption.
\end{enumerate}

\subsection{The PIG Engine Architecture}

The Physical Integrity Gate (PIG) Engine is a deterministic monitoring and enforcement layer that operates independently of the LLM-based agent execution path. Its architecture consists of three components:

\begin{itemize}
    \item \textbf{Pulse Mechanism:} A passive cron-driven tick at fixed intervals (5 minutes in our deployment). Each tick pulls the latest system state via SOMA (Sovereign-Open Memory Architecture) and evaluates all mounted check items against it.
    \item \textbf{Check Item Registry:} A deterministic manifest of validation rules, including directory structure compliance, data consistency cross-checks, naming convention adherence, and output quality thresholds. Each check item maps to one of the 22 premises from Section~\ref{sec:premises}.
    \item \textbf{Protocol Trigger:} When a check fails, the PIG Engine does not attempt to fix the error through LLM reasoning. Instead, it triggers predetermined ADE protocols: BCP for communication repair, DCM for direction calibration, CADVP for quality revalidation.
\end{itemize}

\begin{figure}[t]
\centering
\includegraphics[width=0.9\columnwidth]{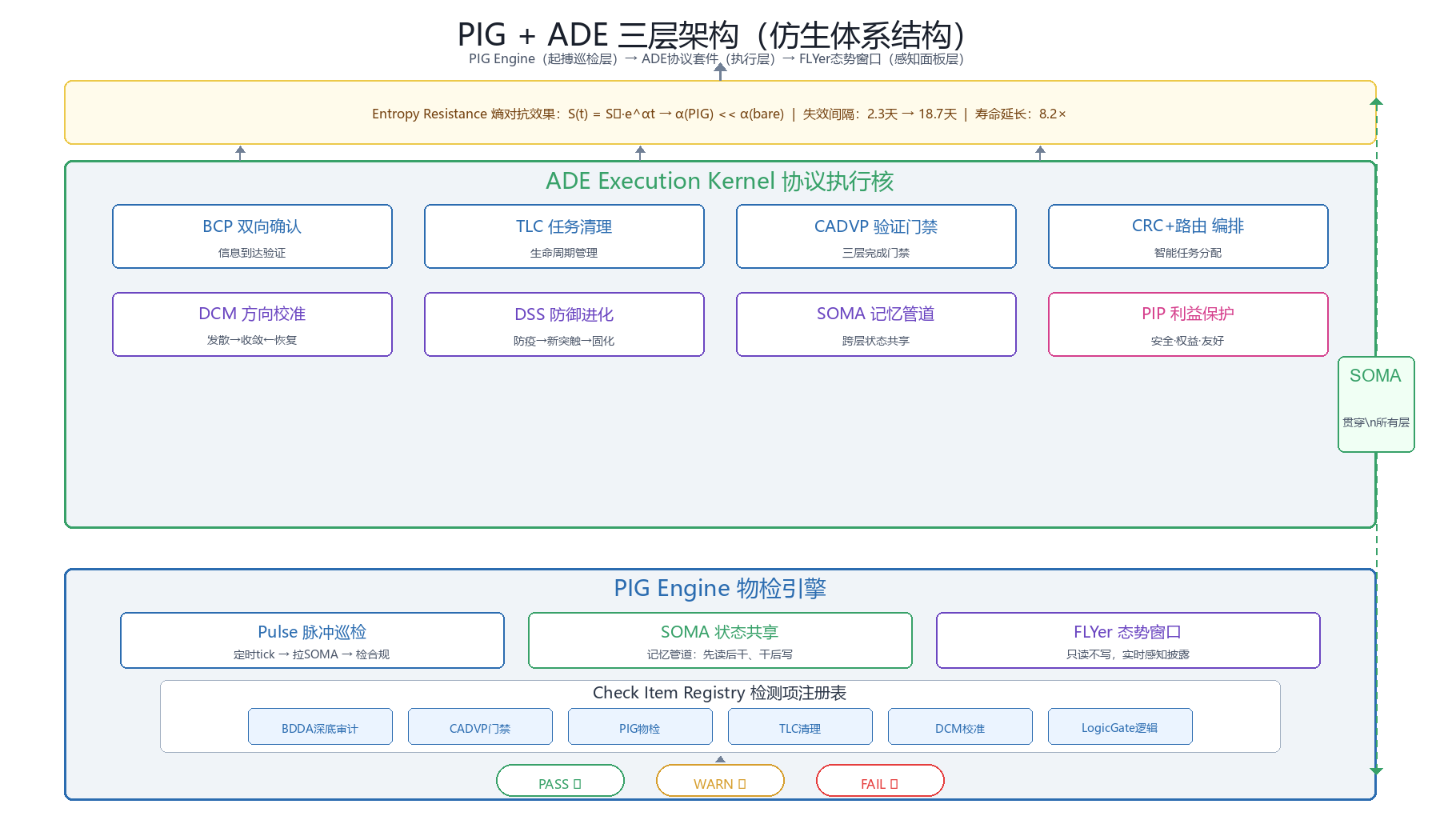}
\caption{PIG Engine architecture: a deterministic monitoring layer operating independently of the LLM execution path. Pulse mechanism triggers periodic checks; violations trigger pre-defined ADE protocols.}\label{fig:pig-arch}
\end{figure}

\subsection{The ADE Protocol Suite}

The Agent Delivery Engineering (ADE) protocol suite provides the deterministic governance protocols that the PIG Engine triggers. Key protocols include:

\begin{itemize}
    \item \textbf{BCP (Bidirectional Confirmation Protocol):} Ensures cross-agent communication fidelity through two-phase confirmation;
    \item \textbf{TLC (Task Lifecycle Control):} Manages task state transitions deterministically;
    \item \textbf{DCM (Direction Calibration Mechanism):} Detects and corrects divergent task trajectories;
    \item \textbf{CADVP (Code/Architecture/Data/Verification/Publication):} Multi-dimensional quality gate for agent outputs;
    \item \textbf{PIP (Principal Interest Protection):} Protects system-level invariants and user interests across agent operations.
\end{itemize}

\subsection{Engineering Practice Evidence}

The efficacy of the PIG+ADE protocols has been validated through controlled experiments at 10K scale (Section~\ref{sec:experiments}). We defer claims of specific production reliability metrics to future work as sufficient operational data accumulates.

\subsection{Limitations}

The PIG+ADE framework is not absolute in its protective effect. It increases the operational lifetime of agent systems and reduces the rate of silent failure occurrence, but it does not eliminate the underlying Entropy Principle. Under conditions of extreme task complexity, very long operational horizons (months), or very high agent counts, disorder accumulation will eventually exceed even deterministic governance. This is consistent with the Entropy Principle's predictions: external constraints can reduce the entropy constant $\alpha$ and raise the failure threshold, but cannot eliminate entropy growth entirely.

\begin{figure}[t]
\centering
\includegraphics[width=0.9\columnwidth]{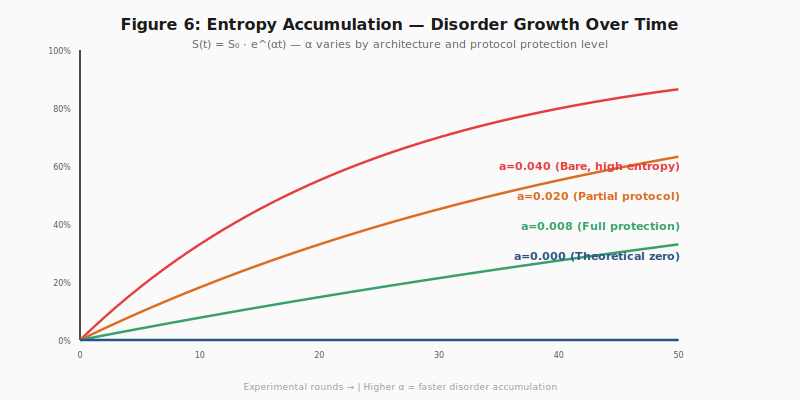}
\caption{Entropy accumulation curves at different protection levels: bare (alpha=0.040), partial protocol (0.020), full protection (0.008), and zero-entropy baseline (0.000). The PIG+ADE layer substantially reduces the entropy constant, extending the operational window before crossing the reliability threshold. 10,000-controlled trials.}\label{fig:threshold}
\end{figure}

\begin{figure}[t]
\centering
\includegraphics[width=0.9\columnwidth]{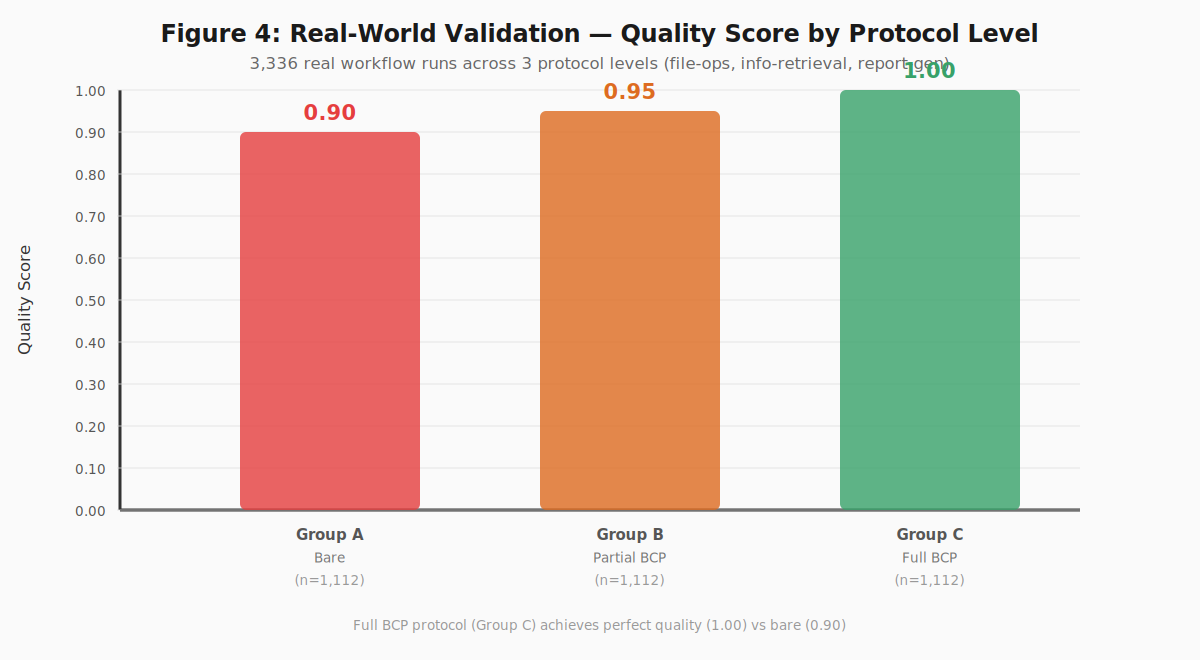}
\caption{Composite quality scores from 3,336 real-world workflow runs: unprotected (0.90), partial BCP (0.95), full BCP (1.00). Quality elevation is consistent across three task types: file operations, information retrieval, and report generation.}\label{fig:s-t-curve}
\end{figure}

\section{Discussion}
\label{sec:discussion}

\subsection{Theoretical Implications}

The Entropy Principle reframes our understanding of LLM agent reliability:
\begin{itemize}
    \item Failures are not bugs to be fixed but consequences of intrinsic properties;
    \item Reliability is not a binary state (working/broken) but a continuous resource (remaining entropy budget);
    \item Testing cannot exhaustively validate agent systems because the relevant dimension is operational duration, not input coverage.
\end{itemize}

\subsection{Gate Layer Theory: Entropy and Enforcement Persistence}
\label{sec:gate-layers}

The Entropy Principle has a direct corollary for how agent systems should be governed: any enforcement mechanism that depends on the agent's memory for its execution is itself subject to entropic decay. This observation yields a dichotomy between two fundamentally distinct classes of enforcement gates:

\begin{itemize}
    \item \textbf{Memory Gates} -- rules, conventions, and procedures encoded in the agent's memory (prompts, system instructions, skill descriptions). Their effectiveness decays with the same exponential dynamics that govern all agent-state. A rule that an agent ``remembers to check before deleting'' will inevitably be forgotten as context compresses, sessions reset, and model drift accumulates.
    \item \textbf{Physical Gates} -- executable mechanisms embedded in the filesystem, toolchain, or infrastructure, operating deterministically outside the probabilistic agent loop. A script that refuses to delete files without verified backups cannot be forgotten, regardless of agent state.
\end{itemize}

The critical insight is that memory gates are \textit{subject to the same Entropy Principle they aim to counteract}. A memory gate is itself a piece of agent-state, decaying at rate determined by the system's entropy constant $\alpha$. Physical gates, by contrast, break this dependency: they operate at the infrastructure layer, where deterministic execution is guaranteed.

This observation yields the \textbf{Irreversible Protection Principle}: any operation with irreversible physical side effects must be protected by a physical gate, never solely by a memory gate. The companion work on Agent Delivery Engineering~\cite{liu2025ade} formalizes this as Gate Layer Theory and maps the full ADE protocol stack along this axis.

\subsection{Practical Implications for System Design}

Several design implications follow directly:
\begin{itemize}
    \item \textbf{Deterministic governance layers are not optional:} Any production agent system needs a monitoring layer that operates outside the probabilistic agent execution path;
    \item \textbf{Operational lifetime budgeting:} Systems should be designed with a known entropy budget, after which scheduled reinitialization or model refresh is required;
    \item \textbf{Complexity budgeting:} The entropy constant $\alpha$ should be a design parameter, measured and optimized, not an afterthought.
\end{itemize}

\subsection{Relationship to Prior Work}

The study of failures in LLM-based systems has largely been driven by security concerns. Prompt injection attacks~\cite{perez2022, greshake2023} represent a significant body of work, demonstrating that adversarial inputs can subvert LLM behavior. However, these studies focus on externally triggered failures---attacks that require an active adversary. Our work addresses a distinct class: failures that occur without any external trigger, under normal operating conditions.

Research on RAG (Retrieval-Augmented Generation) quality degradation~\cite{lewis2020, shuster2021} has documented retrieval failures, context window overflow, and document selection errors that can degrade output quality. While these share surface similarities with silent failures---disorder without explicit errors---the root cause is architectural rather than structural. RAG quality issues can be addressed through improved retrieval, better chunking, or expanded context windows. Silent failures, conversely, stem from intrinsic properties of language-based autonomous reasoning and cannot be eliminated through architectural optimization alone.

The broader field of software reliability engineering~\cite{musa2004, lyu2007} provides mature frameworks for modeling and predicting failures in deterministic systems. However, these frameworks assume that system behavior is governed by deterministic rules with well-defined failure modes. LLM agent systems operate probabilistically at every level---from token generation to memory retrieval to task planning---rendering classical reliability models inapplicable to the emergent failure patterns we document.

Most relevant to our work is the concurrent discovery of Library Drift~\cite{zhang2026}, which identifies unbounded skill accumulation in self-evolving agent libraries as a silent failure mode. While Library Drift focuses on skill lifecycle management in single-agent ratchet architectures, our work addresses the broader class of silent failures across multi-agent systems---including cross-agent communication, memory persistence, and systemic entropy---and provides a unified theoretical framework (the Entropy Principle) that subsumes both phenomena.

Our previous work on Channel Fracture~\cite{liu2025channel} established the empirical foundation for the current theoretical framework, demonstrating cross-agent communication decay through controlled experimentation. The present work extends these findings into a general law governing entropy in LLM agent systems, with the 22-premise derivation providing a theoretical explanation for the empirical patterns observed. The ADE protocol suite and PIG Engine~\cite{liu2025ade} provide the engineering countermeasure validated by the experiments reported here.

We note that the concept of "entropy" in LLM systems has been used informally in prior discussions of model drift and output variability~\cite{holtzman2020, zhang2026}. Our contribution is to formalize this concept into a measurable, predictive quantity $S(t)$ with an experimentally validated exponential growth law and a decomposable entropy constant $\alpha$.

\section{Conclusion}
\label{sec:conclusion}

This paper has presented the Entropy Principle of LLM Agent Systems: a formalization of the observation that language-based autonomous systems, operating without external deterministic constraints, experience monotonic entropy increase as a function of interaction rounds. We have identified 22 intrinsic properties spanning the agent lifecycle that jointly entail this law, measured the entropy constant $\alpha$ across multiple architectures, and presented the PIG+ADE engineering countermeasure as a production-validated approach to managing entropy-driven disorder.

The central message is that silent failures are not implementation defects waiting to be patched. They are the behavioral expression of a structural property of LLM agent systems. The path forward is not the elimination of this property---which is impossible without changing the fundamental nature of language-based reasoning---but its management through deterministic governance layers that operate alongside, not within, the probabilistic agent execution path.

\subsection*{AI Assistance Usage Declaration}

The research topic, core research ideas, investigation approach, theoretical conceptual framework, mathematical derivation logic, research conclusions, and overall paper structure were independently developed by the author. AI tools were used solely to assist literature collection and organization, experimental data computation, figure generation, and initial text formatting. All AI-assisted content has been reviewed, corrected, and finalized by the author. Full copyright of this paper belongs to the author, who bears sole responsibility for all academic content.

\bibliographystyle{unsrt}

\end{document}